\documentclass[a4paper]{jpconf}
\usepackage{graphicx}
\usepackage{amssymb}

\RequirePackage{color}

 \definecolor{MyDarkGreen}{rgb}{0.02,0.60,0.06}

\begin{document}
\title{Maths Meets Myths: Network Investigations of Ancient Narratives}

\author{Ralph~Kenna$^{1,3}$ and P{\'{a}}draig Mac Carron$^2$} 

\address{$^1$ Applied Mathematics Research Centre, Coventry University, CV1 5FB, England}

\address{$^2$ Social and Evolutionary Neuroscience Research Group,
Department of Experimental Psychology,University of Oxford, OX1 3UD, England}

\address{$^3$ The Doctoral College for the Statistical Physics of Complex Systems,
Leipzig-Lorraine-Lviv-Coventry $({\mathbb L}^4)$.}

\ead{r.kenna@coventry.ac.uk; padraig.maccarron@psy.ox.ac.uk}

\begin{abstract}
Three years ago, we initiated a programme of research in which ideas and tools from statistical physics and network theory were applied to the field of comparative mythology.
The eclecticism of the work, together with the  perspectives it delivered, led to widespread media coverage and academic discussion. 
Here we review some aspects of the project, contextualised with a brief history of the long relationship between science and the humanities. 
%sketch the history of applying physics to the humanities and outline some reactions which such excursions can provoke amongst scientists and humanities scholars. 
We focus in particular on an Irish epic, summarising some of the outcomes of our quantitative investigation. 
We also describe the emergence of a new sub-discipline and our hopes for its future. 
%\blue{Finally we present some new results concerning the importance of certain types of genealogical relationships in the Western European epic canon.}
\end{abstract}

%%%%%%%%%%%%%%%%%%%%%%%%%%%%%%%%%%%%%%%%%%%%%%%%%%%%%%%%%%%%%%%%%%%
\section{Introduction}
%%%%%%%%%%%%%%%%%%%%%%%%%%%%%%%%%%%%%%%%%%%%%%%%%%%%%%%%%%%%%%%%%%%
\label{I}
\setcounter{equation}{0}

In 2012 and 2013 we published two papers~\cite{us1,us2} which generated significant amounts of interest within and beyond academia. 
This resulted in invited articles  in science magazines,  newspaper reports, radio interviews and a host of coverage on blogs and online media. 
For us, this was an unprecedented amount of media attention and illustrated the potential that interdisciplinary research has for impact generation. 
So what led to such levels of interest?

Although the distinctions between them are not always sharp, myths differ from legends and folktales. Mythology entails a plethora of characters and timeless narratives outside documented history. 
Legends are couched in a definite historical timeframe and folktales are intentionally fictional. 
Although each of these is broadly accessible, and forms important components of our cultural identities, certain works of mythology are  the  most ethereal. 
Part of the reason for widespread interest in  our work was that it involved the treatment of such ancient corpora in a very modern way -- as complex systems. 

The aim of our research was (and remains) to study networks of characters appearing in different mythologies to facilitate their classification and comparison.
For example, the {\emph{T{\'{a}}in B{\'{o}} C{\'{u}}ailnge}} (Cattle Raid of Cooley) is the most famous epic of Irish mythology. 
The tale describes the invasion of Ulster (the northern part of Ireland) by the armies of queen Medb of Connacht (the western part) and its defence by the warrior C{\'{u}}chulainn. 
Before it was committed to writing by medieval Irish scholars, the {\emph{T{\'{a}}in}} had an extensive oral tradition. 
The historicity of the {\emph{T{\'{a}}in}} has long been  debated. 
Some argue that such narratives corroborate Greek and Roman accounts of the Celts while others object that such tales have no historical basis. 

Medb's royal residence, with king Ailill, was at Cruachan, the ancient capital of Connacht. 
This is identified as a complex of archaeological sites at Rathcroghan (R{\'{a}}th Cruachan, the Ringfort of Cruachan),  %near Tulsk 
in County Roscommon, about 50~km north-west of Athlone. 
In 1864 Samuel Ferguson explored nearby Oweynagat cave and found two Ogham stones there (monoliths inscribed in an early Irish  alphabet).
On returning to Dublin he discovered one letter of his record of these inscriptions to be indistinct. 
He immediately took the night train back to Roscommon and was in the cave by the following noon. 
He translated the inscription as {\tt{FRAICCI MAQI MEDFFI}}, meaning ``Fraic, son of Medf.'' 
One can imagine his excitement, having found what appeared to him to be evidence for a historical Medb
\cite{Waddell}.
Nowadays Medb is believed by some to be an amalgam of several entities and antecedents.
Indeed, in Section~5 we will present evidence which may support this suggestion.

Mythology continues to provoke excitement today. 
%It is familiar, yet intangible. 
Comparative mythology is a well developed subject within the humanities. 
Joseph Campbell maintained that mythological narratives from a variety of
cultures around the world essentially share the same fundamental structure, called the {\emph{monomyth}}
\cite{Campbell}. 
The monomyth is supposed to represent the universality of the human condition; the idea is to explain
and contextualise our world, different societies have constructed different mythologies. 
The  similarities in the structures of these narratives, as Campbell saw it, reflect common or universal human concerns. 
However, the concept of the monomyth has been criticised and mainstream scholarship of comparative mythology values differences as well as similarities~\cite{Consentino,Northup}. 
Nonetheless, the idea of universality (or lack of it) overlaps with a fundamental concept in the statistical physics of critical phenomena. 

Epic mythology also differs from folktales and legend in that it often entails large casts of characters. 
The number of interactions between these characters is vaster still.
We know from statistical physics in general, and sociophysics in particular, the value of treating such medium-to-large data sets as complex systems.
Therefore it is tempting to seek to apply statistical physics techniques -- or at least viewpoints -- to comparative mythology to see what comes out. 
The focus on the collection of interactions in  large sets of characters is a novel approach to comparative mythology and, indeed, to literature generally, because many traditional methods tend to zoom in on the individual. 
We view the method as complementary to traditional approaches -- it certainly cannot replace them, but can perhaps supplement them.

In the next section we briefly outline why physicists like to investigate subjects beyond the physical world and why they should be allowed to do so. In Section~~\ref{III} we provide the context for our work by summarising some aspects of Irish mythology.
In Section~~\ref{IV} we introduce our methodology and we apply it to three iconic European epics in Section~~\ref{V}. We conclude in Section~\ref{VI}.

%%%%%%%%%%%%%%%%%%%%%%%%%%%%%%%%%%%%%%%%%%%%%%%%%%%%%%%%%%%%%%%%%%%
\section{Why do Scientists want to do Research in the Humanities?}
%%%%%%%%%%%%%%%%%%%%%%%%%%%%%%%%%%%%%%%%%%%%%%%%%%%%%%%%%%%%%%%%%%%
\label{II}
\setcounter{equation}{0}

Why do physicists wish to delve into subject areas covered traditionally by other disciplines and in which they have little or no expertise?
Physicists, and statistical physicists in particular, are interested in {\emph{ complex systems}} -- how 
properties can emerge in an aggregate system which are absent from the microscopic agents which comprise it.
The classic example from a statistical physicist's point of view is phase transitions; 
water turns to ice or steam at sufficiently low or high temperatures. 
Yet its constituent parts, the molecules of H$_2$O from which it is constructed, do not themselves manifest  different phases.
Another example is that of flowing water.
It can exhibit turbulence, but an individual molecule cannot.
Similarly, the characteristics of a population are not simple aggregates of the behaviours of individuals. Examples include flocking,  language, culture, and mythology itself.
Each of these is a feature of the collective rather than the individual.
In recent decades, simple models, many inspired by statistical physics, have been developed to account for such social phenomena. 
Disciplines with names such as {\emph{complexity science}}, {\emph{sociophysics}} and {\emph{econophysics}} have emerged and are hotbeds of activity. 

Thus physicists are not only interested in the physical and, indeed, physics does not only concern the physical. 
There are many physicists and their motivations are varied. 
Some are interested in the practical and that which can be applied.
Others are motivated by the challenge of tackling problems which are difficult intellectually.
Still others are driven by curiosity. 
It is often  the latter type  that  engages in cross-disciplinary pursuits.
With curiosity as a map in one hand, and holding the passport that is academic freedom in the other, curious physicists  are happy to make forays into anything that interests them.

Some of this activity is interdisciplinary involving cooperation with sociologists, psychologists and economists as well as humanities scholars.
The relationship between science and humanities has sometimes been perceived to be a difficult one.
{\emph{The Two Cultures}} was the title of part of an influential Lecture by the English physical chemist and novelist Charles Snow in 1959~\cite{Snow}. 
The gist of his talk was that our society, its intellectual life, priorities, and systems of education, is split into two cultures -- humanities and the arts on one hand, and science on the other. 
Snow argued that difficulties in communication between the two  is a significant impediment to solving many of the world's problems. 
Examples of these problems include 
informing and influencing politicians,  
developing  science and technology strategies,
deciding policies for the future of education 
and issues of cultural fragmentation.
Snow's diagnosis provoked an enormous amount of controversy and is still debated to this day. 

%Mistrust between the sciences and the humanities, however, goes back further. 
%In the early part of the 19th Century, John Keats argued, in his poem {\emph{Lamia}} that
%the then new sciences of physics and chemistry might ``unweave the rainbow.''
%In other words, Keats was lamenting that science would unravel some of the great beauties and mysteries of nature. 
%In the 20th Century Irving Babbitt,  the  American academic and literary critic
% ``The humanities need to be defended today against the encroachments of physical science.''

The {\emph{Two Cultures}} dichotomy is unfortunate in our view, as science and the humanities are perhaps not so far apart as it pretends.
Indeed, the spirit of curiosity that drives many physicists is the same one that motivates many activities in the humanities. 
These are often not financially-driven or even applications-driven pursuits. 
Such research activity  may be  removed from everyday life but, in pushing at theoretical and  methodological  boundaries, they can launch new and unpredictable developments.
Statistical physics and the humanities have a common cousin in the social sciences. 
In the 18th century people noticed regularities in numbers of events such births, deaths, etc.
This was surprising because individuals themselves are unpredictable.
This partly motivated the development of  statistical approaches to  many-body systems
\cite{Fortunato}.

The word {\emph{sociophysics}} is related to the older term {\emph{social physics}}.
Both refer to the study of social systems from the perspective of physics or using concepts or tools of physics. 
The term {\emph{sociology}} itself was first used by Auguste Comte, a founder of the discipline, as an alternative to the term social physics. 
It is interesting to compare the following two quotes in this context:
\begin{quote}
``Now that the human mind has grasped celestial and terrestrial physics,
 --- mechanical and chemical; organic physics, both vegetable and animal, ---
 there remains one science to fill up the series of sciences of observation, 
--  social physics. This is what men have now most need of.''
%``Now that the human mind has grasped celestial and terrestrial physics, …there remains one science … social physics. This is what men have now most need of.''
\end{quote}
\begin{quote}
``I think the next century will be the century of complexity.
We have already discovered the basic laws that govern matter and understand all the normal situations. We don't know how the laws fit together, and what happens under extreme conditions.''
% But I expect we will find a complete unified theory sometime this century. The is no limit to the complexity that we can build using those basic laws.''
\end{quote}
The first of these is from  Compte in the 1850's~\cite{Compte}. 
The second is atributed to Steven Hawking in 2000~\cite{Hawking}.
%Thus, although separated by  almost 150 years, the application of physics to social and/or complex remains seen as a ``next big thing.''
Given the long-standing links between statistical physics and sociology, and the current interest in complexity, it is perhaps not such a great leap to investigate aspects of humanities (literature, mythology) which may be viewed as many-body, interacting systems.

\section{Aspects of Irish Epic Mythology}
%%%%%%%%%%%%%%%%%%%%%%%%%%%%%%%%%%%%%%%%%%%%%%%%%%%%%%%%%%%%%%%%%%%
\label{III}
\setcounter{equation}{0}

%Rathcroghan (R{\'{a}}th Cruachan, the Ringfort of Cruachan),  is a complex of ancient sites near Tulsk in County Roscommon and has been identified as the traditional capital of Connacht.
%It features prominently in the Ulster Cycle, one of the four great groupings of Irish mythology, where it is described as
Besides being the location of the opening scenes of the  {\emph{T{\'{a}}in B{\'{o}} C{\'{u}}ailnge}}, Cruachan features in {\emph{T{\'{a}}in B{\'{o}} Flidhais}} and {\emph{T{\'{a}}in B{\'{o}} Aingen}} (also known as {\emph{Echtra Nera{\'{i}}}}). 
Through Oweynagat, it is also associated with {\emph{Samhain}}, the Gaelic seasonal festival which gave rise to the modern Halloween. 
Archaeological evidence supports the suggestion that the site was an important location for ritualistic gatherings and a significant cemetery, although there is little archaeological evidence to support the assertion that it was a royal residence as often described in the mythological literature~\cite{Waddell,WaFe09}.

To perform a broad investigation from  a non-material perspective, we used social-network analysis.
% (Wasserman and Faust, 1994). 
% The application of mathematical and physics techniques to social systems has emerged in recent years as a sub-discipline called sociophysics (Galam, 2004). 
% In an interdisciplinary project, these tools have recently been applied to study and compare the social networks which underlie the societies depicted in mythological narratives (Mac Carron and Kenna, 2012). 
Here we report on these investigations, with a particular focus on the {\emph{T{\'{a}}in B{\'{o}} C{\'{u}}ailnge}}.
This belongs to the Ulster Cycle,   one of the four great groupings of Irish mythology.
Early Irish literature presents ideal material for a social-network approach because of the preponderance of characters they contain, as well as the interactions between them. 
The publisher and Celticist Alfred Nutt estimated that known literature of the Ulster Cycle alone would occupy at least 2,000 pages of a modern volume if all repetitions were edited out~\cite{Nutt,Dunn}.
The question we asked, is how these mythological networks compare to other social networks; what are their universal features, if any, and what are the distinguishing ones?
%If the social networks depicted in ancient texts appeared realistic, we surmised, perhaps they may reflect some degree of historical reality.

We started with the {\emph{T{\'{a}}in B{\'{o}} C{\'{u}}ailnge}}  because it is the most famous epic of Irish mythology,  frequently compared to Homer's {\emph{Iliad}} and  the Anglo-Saxon {\emph{Beowulf.}} 
%The tale describes the invasion of Ulster by the armies of Queen Medb of Connacht and the defence by C{\'{u}}chulainn, Ireland's most famous hero. 
Related to the {\emph{T{\'{a}}in}} itself are a number of pre-tales and tangential tales (remsc{\'{e}}la) which give the backgrounds and exploits of the main characters. 
The {\emph{T{\'{a}}in}} has come down to us in three recensions. 
The first has been reconstructed from partial texts contained in {\emph{Lebor na hUidre}} (the Book of the Dun Cow, dating from the 11th or 12th century and compiled at the monastery at Clonmacnoise, 20 km south of Athlone) and {\emph{Lebor Buide Lec{\'{a}}in}} (the Yellow Book of Lecan, a 14th century manuscript) and other sources. 
The second, later recension is found in {\emph{Lebor Laignech}} (the {{Book of Leinster}}, a 12th century manuscript formerly known as the {\emph{Lebor na Nuachongb{\'{a}}la}} or {{Book of Nuachongb{\'{a}}il}}, after a monastic site at Oughaval in County Laois). A third recension comes from fragments of later manuscripts and is incomplete.
Two popular English translations of the {\emph{T{\'{a}}in}}~\cite{Kinsella,Carson} are mainly based on the first recension, although they each include some passages from the second. We base our analysis on~\cite{Kinsella} which therefore, for the purposes of this analysis, serves as a proxy for what is commonly understood as the {\emph{T{\'{a}}in B{\'{o}} C{\'{u}}ailnge}}.

Before they were committed to writing, the {\emph{T{\'{a}}in}} and other mythological narratives had extensive oral histories. 
The {\emph{T{\'{a}}in}} was dated by medieval Irish scholars to the first centuries BC but this may have been an attempt by Christian monks to artificially synchronise oral traditions with biblical and classical history. 
It has been speculated that Christian scribes may have sought to de-deify Pagan gods to the level of (super-) human heroes to dilute Pagan influences~\cite{Green}. 
Similar reasons may have led to Rathcroghan mound, for example, being depicted as a residence rather than a religious site~\cite{Wa12}. 
It has been argued that heroic literature, written down in Christian times, contains attempts to fit to contemporary notions in order to make it more acceptable to Christian society~\cite{Be98,Wa11}. 
Indeed, the {{Book of Leinster}} version of the {\emph{T{\'{a}}in}} ends with scribal notes in Latin and Irish which state~\cite{Dunn} 

\begin{quote}``A blessing be upon all such as shall faithfully keep the {\emph{T{\'{a}}in}} in memory as it stands here and shall not add any other form to it.

``I, however, who have copied this history, or more truly legend, give no credence to various incidents narrated in this history or legend. For, some things herein are the feats of jugglery of demons, sundry others poetic figments, a few are probable, others improbable, and even more invented for the delectation of fools.''
\end{quote}

Undoubtedly much material, which has come down to us today as mythology, has been influenced by religion and politics. 
Jeffrey Gantz  has described the early Irish tales as verging ``upon wishful thinking, if not outright propaganda''~\cite{Gantz}. 
For example, the {\emph{Dindshenchas}} texts on Tara (onomastic texts giving the origins of place-names and traditions associated with places) are believed to have been compiled to bolster the claims of the southern U{\'{i}} N{\'{e}}ill (from the middle Kingdom of Mide) over those of Brian Boru in Munster (Ireland's southern province)~\cite{Wa05}. 
Myth could be used to explain why things are as they are and to justify the ambitions of dynasties~\cite{Byrne}. 
The {\emph{Dindshenchas}} and other texts in turn influenced Geoffrey Keating's famous seventeenth-century {\emph{History of Ireland}~\cite{Cunningham}. In the words of Keating (referring to the later Fenian Cycle of Irish mythology), 
\begin{quote}
``And whoever should say that Fionn and the Fian never existed, would not be stating truth... For it has been delivered to us from mouth to mouth that Fionn and the Fian existed; and, moreover, there are numerous documents that testify to this.''
\end{quote}
%(Fionn and the Fian referred to here feature in the stories of the Fenian Cycle of Irish mythology.)

Through the 18th century, the historicity of Irish mythology was hotly debated. 
On the one hand, writers such as Edward Ledwich rejected the idea of a sophisticated civilization in ancient Ireland; on the other hand, Charles O'Connor and Sylvester O'Halloran defended the notion of a heroic past~\cite{Wa05}. 
The historicity of the {\emph{T{\'{a}}in}} continued to be contested in the 20th century. 
For example, while Thomas O'Rahilly~\cite{Rah} objected that such tales have no historical basis whatsoever, Myles Dillon~\cite{Dil} believed that heroic sagas give ``a picture of pre-Christian Ireland which seems genuine.'' 
Kenneth Jackson~\cite{Jac} argued that while such narratives corroborate Greek and Roman accounts of the Celts and offer us a ``window on the Iron Age'', ``the characters Conchobar and C{\'{u}}chulainn, Ailill and Medb and the rest, and the events of the Cattle Raid of Cooley, are themselves entirely legendary and purely un-historical. 
But this does not mean that the traditional background, the setting, in which the Ulster cycle was built up is bogus.'' 
Gantz~\cite{Gantz} concludes that ``the Irish society represented by the Ulster Cycle is ... very similar to the Gaulish civilization depicted by Caesar.'' 
So may such old myths contain kernels of truth or are they entirely fanciful? 

Currently there remains little direct material evidence for the events and the society in the {\emph{T{\'{a}}in}}. However, there is also ``no definite evidence against the medieval Irish belief that Conchobar and his court flourished around the time of Christ''~\cite{Byrne}. 
Myles Dillon and Nora Chadwick~\cite{DiCh67} have argued that while the {\emph{Lebor Gab{\'{a}}la {\'{E}}renn}}  ({{Book of Invasions}}) which forms a major part of the Mythological Cycle, has no value as history, the {\emph{T{\'{a}}in}} gives a more authentic picture of Iron-age Irish society. 
Using material from annals and genealogies to interpret ancient Irish epics, Francis Byrne examined the political landscape of early Irish kings and high-kings and suggested that the wars between Connacht and Ulster may indeed have occurred at any time before the fifth century. 
He suggests that such heroic tales usually contain ``a kernel of historical fact''~\cite{Byrne}. 
There is also some evidence that characteristics of the landscape are reflected in the story~\cite{Lynn}. 
Thus, the extent to which Irish mythological narratives may or may not reflect traces of a historical, albeit distorted, version of an ancient reality continues to fascinate and generates an enormous amount of public interest~\cite{Green,Edel}. 
%For this reason, we considered it worthwhile to examine what modern mathematical techniques  have to say on the matter. 

There are obvious caveats and limitations regarding the material to be analysed. 
These include enormous political and religious influences as discussed above, the obvious lapses in time between the creations of the medieval tracts and the societies and events they purport to describe and the use of modern texts (in English) as proxies for the originals. 
The reliability of any statistical analysis is dependent on the sample size, the larger the better. 
Ideally we would prefer extensive narratives with a large numbers of characters and plenty of interactions between them for increased statistical accuracy. 
Nonetheless, we can only analyse the material which has come down to us and we can only apply the techniques which we possess. 
The reader should keep in mind that we cannot offer proofs – only evidence and comparisons from new perspectives. 
It is up to the experts in archaeology and mythology to inform further and ultimately it is up to the readers to make informed judgements of their own. 
Our purpose is to help inform such judgements from a wholly new perspective.

%We are conscious that many readers come to this exposition with an interest in antiquities in general and in Rathcroghan in particular. While possibly knowledgeable in mythology or archaeology, we cannot expect the reader to have a theoretical physics background or to be versed in the new sub-fields of sociophysics and complexity theory. For the benefit of these readers, we provide a brief picture of these fields and of complex networks in general – from a non-mathematical point of view.  After these, we report on the application of these techniques to the analysis of the social network underlying {\emph{T{\'{a}}in B{\'{o}} C{\'{u}}ailnge}}.

%%%%%%%%%%%%%%%%%%%%%%%%%%%%%%%%%%%%%%%%%%%%%%%%%%%%%%%%%%%%%%%%%%%
\section{Complex Networks}
%%%%%%%%%%%%%%%%%%%%%%%%%%%%%%%%%%%%%%%%%%%%%%%%%%%%%%%%%%%%%%%%%%%
\label{IV}
\setcounter{equation}{0}

Networks can be used to quantitatively characterise a large range of structures, such as the internet, power grids, transport networks  and food-webs. 
In a social network, the nodes are representations of people and the links represent interactions or acquaintanceships between those people. 
A number of statistical tools have been introduced to describe and quantify various properties of networks. 
Here we briefly describe some of these tools in a non-technical manner, referring the reader to~\cite{us1} for details.

There is a famous notion in sociology of {\emph{six degrees of separation}}. 
This is the idea that, despite the world’s population of about seven billion, everyone is, on average, only about six steps away from any other person. 
In this case the number six is an example of the notion of average path length. 
This is a measure of the connectivity of the network and tends to be very short in social networks. 
A second mathematical measure of connectivity is the {\emph{clustering coefficient}}, which gives an indication as to how cliqued a network is. 
In social networks the clustering coefficient tends therefore to be very high compared, for example, to a random network. 
These measures of connectivity illustrate that social networks have different characteristics to other types of complex networks. 
The highly connected nature of social networks is a property which is known as {\emph{small world}}~\cite{WS}.

The {\emph{degree}}  of a node in a network is the number of links it has. 
In the narratives we  deal with, the degree of a character is the number of other characters he or she interacts with. 
The degree distribution in social networks  tends to be well approximated by a power law and is often found to be scale free~\cite{Amaral}. 
Removing the most connected characters in networks with scale-free degree distributions causes them to break down and become disconnected. 
However, removing characters at random does not strongly affect networks as they are robust to random attacks but not to targeted attacks.

In real-life social networks, people tend to be friends with other people who are similar to themselves. 
For example, popular people are more likely to have popular friends. 
We say that a network in which most nodes have this property is {\emph{assortative}}~\cite{Newman2002}. 
Networks that do not have this property are {\emph{disassortative.}} 
It turns out that assortativity is a very important property which distinguishes social networks from other networks. 
To investigate the assortativity properties of a network, one simply compares the degree of a node (acquaintances of a character) to the degrees of its neighbours. 

In the past decades, a large array of social networks has been studied.
Examples include networks of company directors~\cite{Davis}, jazz musicians~\cite{GleiserDanon}, movie actors~\cite{Amaral}, users of online-forums~\cite{Kujawski}, and scientific co-authors~\cite{Newman2001}. 
Catalogues of such social networks have been assembled and they are normally small world, highly clustered, assortative and scale free~\cite{NewmanPark2003}. 
Thus the general properties of real social networks are well established and well documented.
At the other extreme, physicists have also examined the social networks underlying the panoply of characters that have appeared down the years in Marvel comics~\cite{Alberich,Gleiser}. 
Analysis of the network found that it does not have all the features of a real social network.
Although it was found to be highly connected, it is also highly disassortative. 
All of the very high degree characters were too highly connected, making the network appear artificial.
In the academic literature on networks, there are few if any other analyses of artificial or fictional networks. 

For this reason, we also constructed and analysed the networks of several works of fiction. 
%We compiled network data for Harry Potter as representing an obviously fictional text. 
%We also examined The Fellowship of the Ring (from Lord of the Rings) because it is a fictional work which was inspired by mythology. 
Clearly a small sample is not meant to embody all of the network features of the entire corpus of world fiction. 
Rather, they are examples of clear works of fiction --- fiction involving fabulous characters, which do not attempt to emulate reality. 
%In addition, we analysed Shakespeare's Richard III and Hugo's Les Mis{\'{e}}rables (because data were readily available) and the Marvel Universe. 
It is hoped that, in the future, other researchers will analyse other works of fiction, so that a more extensive list of typical features will be built.
For the moment, we found that each of the fictional networks we analysed were small world, just like real social networks. 
However, none of their degree distributions appeared to follow a clear power law and  they were all disassortative. 
Moreover, unlike real social networks, the fictional ones were very robust upon targeted removal of the most connected nodes. 
%To summarise, their degree distributions, disassortativity and robustness distinguish the fictional social networks which we examined from real ones.  

%%%%%%%%%%%%%%%%%%%%%%%%%%%%%%%%%%%%%%%%%%%%%%%%%%%%%%%%%%%%%%%%%%%
\section{Network Comparison of the {\emph{T{\'{a}}in B{\'{o}} C{\'{u}}ailnge}}, the {\emph{Iliad}} and {\emph{Beowulf}}}
%%%%%%%%%%%%%%%%%%%%%%%%%%%%%%%%%%%%%%%%%%%%%%%%%%%%%%%%%%%%%%%%%%%
\label{V}
\setcounter{equation}{0}

The {\emph{T{\'{a}}in B{\'{o}} C{\'{u}}ailnge}} is frequently compared to other European epics, in particular to {\emph{Beowulf}} and the {\emph{Iliad}}. 
{\emph{Beowulf}} is an Old English heroic epic, set in Scandinavia. 
A single codex survives which has been estimated to date from between the 8th and 11th centuries. 
The story relates the coming of Beowulf, a Gaetish hero, to the assistance of Hrothgar, king of the Danes. 
After slaying two monsters, Beowulf returns to Sweden to become king of the Geats and, following another fabulous encounter many years later, is fatally wounded. 
Although the poem is embellished by obvious fictional elements, such as monsters and a dragon, archaeological excavations in Denmark and Sweden offer support for the historicity associated with some of the human characters~\cite{Anderson1999}. 
Nonetheless, the character Beowulf himself is mostly believed not to have existed in reality 
\cite{Klaeber1950,Chambers1959}. 

The  {\emph{Iliad}} is an epic poem attributed to Homer and is dated to the 8th century BC. 
It is set during the final year of the war between the Trojans and a coalition of besieging Greek forces. 
It relates a quarrel between Agamemnon, king of Mycenae and leader of the Greeks, and Achilles, their greatest hero and counterpart to C{\'{u}}úchulainn.  
Also much debated throughout the years~\cite{Wood1998}, some historians and archaeologists maintained that the {\emph{Iliad}} is entirely fictional~\cite{Finley1954}, while recent evidence suggests that the story may be based on a historical conflict around the 12th century BC interwoven with elements of fiction~\cite{Kraft2003,Korfmann2004,Papamarinopoulos2012}.

%...........................................................................
\begin{figure*}[t]
\begin{center}
\includegraphics[width=1.0\columnwidth,angle=0]{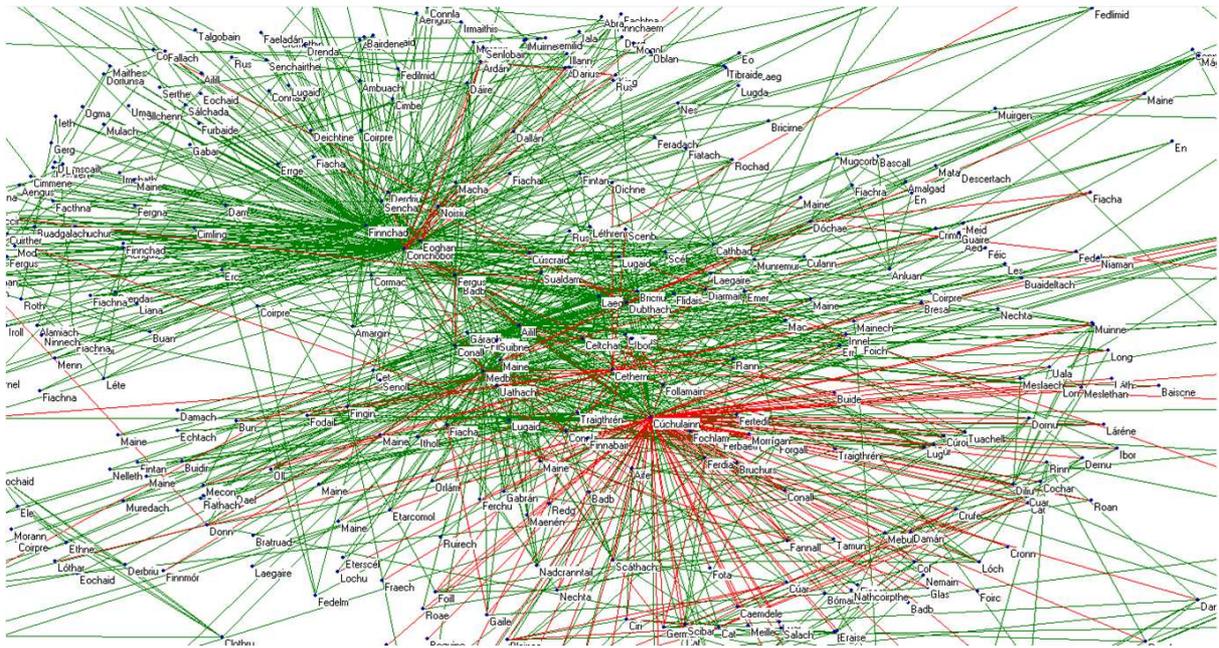}
\label{fig1}
\end{center}
\vspace{0.3cm}
\caption{The {\emph{T{\'{a}}in}}  network. Green links are positive (``friendly'') and red ones are negative (``hostile''). 
Two main  clusters are identifiable by eye. 
The upper left is that of Ulster, with Conchobor at its centre.
The lower right is that of Connacht, centred on Medb and Ailill. 
Between them is Fergus mac Roich and other warriors exiled from Ulster.
C{\'{u}}chulainn himself, although an Ulster warrior, is embedded amongst the Connacht foe as the majority of his links are conflictual.
}
\end{figure*}
%..........................................................................

 The data for the networks were harvested by carefully reading each of the three narratives and entering all characters' names into databases, meticulously listing the characters they interact with. 
We defined two distinct types of interactions: links were designated as {\emph{positive}} or ``friendly'' when two characters know each other, are related, speak to one another, or appear in a small congregation together. 
Links were deemed {\emph{negative}} or ``hostile'' if two characters meet in combat. 
We also assigned a weight to the links between characters, based on how often they meet each other. %For example, in the {\emph{T{\'{a}}in}}, king Ailill and queen Medb of Connacht are more strongly linked than Ailill is linked to C{\'{u}}chulainn. 
In total, Kinsella's {\emph{T{\'{a}}in}} has 404 characters and 1,233 links. 
{\emph{Beowulf}} and the {\emph{Iliad}} were found to have 74 and 716 unique characters and 165 and 2,650 links between them, respectively.
The {\emph{T{\'{a}}in}} network is depicted in Fig.\ref{fig1}.

% Visual inspection of the network of friendly links allows one to identify obvious communities (Figure 1). 
% This is because in the {\emph{T{\'{a}}in}}, for example, characters from Ulster more commonly associate with other characters from Ulster, and characters from Connacht associate with characters from Connacht. 
% The opposite occurs for hostile networks as opposing sides are linked through conflict.

All three epics were found to be small world; they have low average path lengths and high clustering coefficients. 
All the networks became disconnected when the most highly linked 5\% of characters were removed in a targeted manner. But they were all robust when 5\% of characters were removed at random. 
These are characteristics of real social networks. 
However, of the three narratives, only the {\emph{Iliad}} turned out to be clearly assortative. 
{\emph{Beowulf}} was mildly disassortative and the {\emph{T{\'{a}}in}} was highly disassortative. 
%This means only the {\emph{Iliad}} looks realistic. 
%{\emph{Beowulf}} appears similar to the fictional narratives we examined and the {\emph{T{\'{a}}in} more strongly so.

Thus the social-network properties of the {\emph{Iliad}} are similar to those of real social network and dissimilar to the fictitious social networks listed above. 
This is what one might expect if the narrative is a reasonably accurate portrayal of a real society. 
Obviously it does not categorically prove that the {\emph{Iliad}} is based on real events, but it may be interpreted as evidence to support the case for some degree of historicity~\cite{Kraft2003,Korfmann2004,Papamarinopoulos2012}. 

By the same token, the disassortativity of {\emph{Beowulf}} may be a signal of its artificiality. 
However, recall that archaeological evidence suggests that the society in {\emph{Beowulf}} may be based on reality but the eponymous character is believed fictitious. 
This clue prompted us to remove the character Beowulf from the network and to reanalyse it. Interestingly, the resulting network is assortative!

Thus, if assortativity and disassortativity are good markers of reality (or at least realism) and fiction respectively, our analyses of the {\emph{Iliad}} and {\emph{Beowulf}} and the {\emph{T{\'{a}}in}} reflect some of the recent archaeological evidence and opinion; 
the {\emph{Iliad}} appears realistic. 
{\emph{Beowulf}} appears fictional at first sight, but on closer inspection, the artificiality is seen to reside in the lead character (as well as in other, more obvious, fabulous elements); 
{\emph{Beowulf}} the society, without Beowulf the character, appears realistic. 
However the social network in the {\emph{T{\'{a}}in}} is strongly disassortative. 
By the above criteria, it appears fictional. 
We ask wherein resides the artificiality of the {\emph{T{\'{a}}in}}.
Does the entire society appear fictional or is its artificiality localised in a few characters, as {\emph{Beowulf}}'s artificiality is localised in Beowulf himself?

To investigate further, we next turned our attention to the degree distributions of the networks. 
We found that the distributions for all three epics are compatible with power laws, just like real social networks. 
Moreover, there is a striking similarity between the degree distributions of {\emph{Beowulf}} and the {\emph{T{\'{a}}in}} for all but the top six most connected characters in the larger set -- see Fig.2(a). 
The solid line in Fig.2(a) is a curve which is best fitted to the data points which correspond to {\emph{Beowulf}}. 
The six rightmost data points, which represent the six most connected characters in the {\emph{T{\'{a}}in}}, are offset relative to the solid {\emph{Beowulf}} line. 
This means that the degrees of these characters are too large relative to the characters of {\emph{Beowulf}}, and, indeed, relative to the other characters in the {\emph{T{\'{a}}in}}. 
The dashed line is a “best fit” to the {\emph{T{\'{a}}in}} data. 
The anomalous six data points on the right pull the line upwards in that region of the plot. 
This hints that the network structure of the {\emph{T{\'{a}}in}} may be similar to that of {\emph{Beowulf}}, but for the six anomalous characters in the Irish narrative. 
This therefore might be an indication of where the artificiality of the {\emph{T{\'{a}}in}}, is located. 
		
The six anomalous characters in the {\emph{T{\'{a}}in}} network are Conchobor mac Nessa, C{\'{u}}chulainn, Finnchad Fer Benn (Conchobor' s son), Ailill mac M{\'{a}}gach, Medb and Fergus mac Roich. 
Four of these are those identified by Jackson as unhistorical~\cite{Jac}.
To match the  {\emph{T{\'{a}}in}} line with the  {\emph{Beowulf}} one, the degrees of these six characters would have to be reduced. 
In terms of the narrative, this means that the number of interactions they have with other characters would have to be lowered. 
To do this in a systematic manner, we define a weak link as one that occurs when two characters meet only once in the entire narrative. 
We then speculate that some such weak links are proxy interactions in the tale. 
For example, the single encounter between Medb and one of her warriors chosen to fight C{\'{u}}chulainn may rather represent an encounter between Medb's proxy and that individual. 
Removing the weak links from the top six {\emph{T{\'{a}}in}} characters reduces their respective degrees. 
Note what we have done here is  removed weak links – not removed the characters themselves!
Fig.2(b) shows the degree distributions for the adjusted  networks. 
Now the top six points of the {\emph{T{\'{a}}in}} no longer offset the best-fit line and  the data points fall closer to this line for the high degree characters. 

We next re-examined the assortativity of the {\emph{T{\'{a}}in}}. 
Indeed, the removal of the weakest links of the six anomalous characters has the effect of rendering the entire network assortative. 
This supports the suggestion that the apparent artificiality of the {\emph{T{\'{a}}in}} network is associated with just these top six characters. 
It also supports the notion that these characters may each be amalgams of several entities, as suggested in the Introduction.
Now all of the {\emph{T{\'{a}}in}} network properties which we investigated are similar to those of real social networks. 
We may conclude that, on the basis of network analysis alone, the adjusted {\emph{T{\'{a}}in}} network is about as realistic (assortative) as that of {\emph{Beowulf}}.

%...........................................................................
\begin{figure*}[t]
\begin{center}
\includegraphics[width=0.47\columnwidth,angle=0]{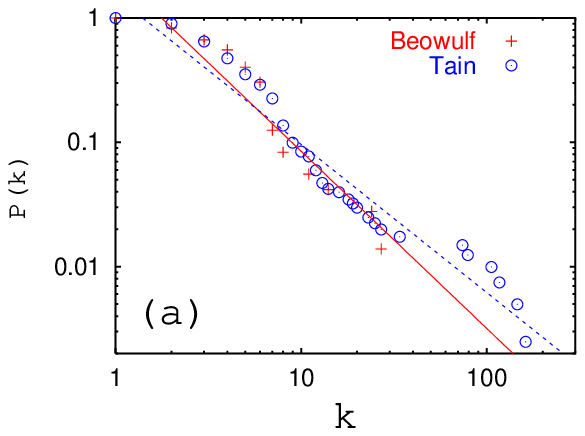}
\includegraphics[width=0.47\columnwidth,angle=0]{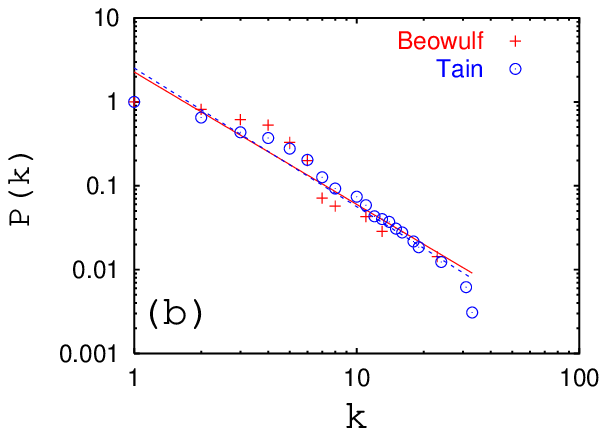}
\label{fig2}
\end{center}
\vspace{0.3cm}
\caption{(a) the degree distributions for {\emph{Beowulf}} and the {\emph{T{\'{a}}in}} overlap except for the six most highly connected characters in the Irish epic.
(b) Removing the weakest links of those characters renders the network similar to that of {\emph{Beowulf}} with the eponymous protagonist removed. Both adjusted networks are assortative.}
\end{figure*}
%..........................................................................

%%%%%%%%%%%%%%%%%%%%%%%%%%%%%%%%%%%%%%%%%%%%%%%%%%%%%%%%%%%%%%%%%%%
\section{Conclusions}
%%%%%%%%%%%%%%%%%%%%%%%%%%%%%%%%%%%%%%%%%%%%%%%%%%%%%%%%%%%%%%%%%%%
\label{VI}
\setcounter{equation}{0}

It has been claimed that archaeological evidence supports the historicity of some of the events of the {\emph{Iliad}}.  
Manfred Korfmann argues that the onus lies now with those who deny any historical association between events in Late Bronze Age Troy and the {\emph{Iliad}} to defend their positions, rather than the other way around~\cite{Korfmann2004}. 
There is also material evidence for the existence of some of the characters of {\emph{Beowulf}}. However, the eponymous protagonist is not believed to have been a real person. 
There is little material evidence to support the historicity of the society or events depicted in the {\emph{T{\'{a}}in B{\'{o}} C{\'{u}}ailnge}}, but there is no definite evidence to the contrary either~\cite{Byrne}.  

While we cannot comment on the purported events described in these narratives, here we have used ideas from modern statistical physics to analyse the relationships that underpin the societies depicted in them. 
We can draw on the experience and knowledge garnered in the complex-network and sociophysics community over recent years. 
This indicates, in particular, that real social networks have certain characteristics and assortativity appears to be most important amongst these.

%, including assortativity.
%In a first attempt to ascertain whether assortativity may be a marker for reality of social networks, we analysed a number of fictional narratives, some of which are entirely fanciful. 
%We found that indeed, our first foray into the topic indicates that disassortativity may be a marker for artificiality. 

On the spectrum from the assortative to the disassortative, the {\emph{Iliad}} has properties most similar to those of real social networks, {\emph{Beowulf}} is mildly disassortative and the {\emph{T{\'{a}}in}} is strongly so. 
The beauty of applying maths to myths 
%mathematics to mythology  approach 
is that, unlike material artefacts, we can adjust the data set to address the question of what it would require to render these networks realistic. 
In the case of {\emph{Beowulf}}, we follow established opinion and remove the protagonist to find the remaining network assortative. 
Thus, while the entire network is not realistic, we may suggest that an assortative subset has properties akin to real social networks, and this subset has corroborative evidence of historicity.

For the {\emph{T{\'{a}}in}}, the social network of the full narrative is disassortative.
%initially seems similar to that of the Marvel Universe, perhaps indicating it is the Iron Age equivalent of a comic book.  
On this basis, the society it describes appears to have ``no credence'' as the scribe noted at the end of the Book of Leinster. 
However, from a network-theoretic perspective,  the artificiality is related to the top six most connected characters. 
Similar to the superheroes of the Marvel Universe, for example, they are too well connected. 
Speculation that these characters may be based on amalgams of a number of entities and proxies, suggests removal of their associated weak social ties. 
Indeed, the resulting network is assortative, similar to the {\emph{Iliad}} and to other real social networks and very different to the society of the Marvel Universe and other works of intended fiction we examined. 
On this purely theoretical basis, we suggest that if the society in the {\emph{T{\'{a}}in}} is to be believed, each of the top six characters is likely to have been formed by amalgamations of characters, as the narrative was passed orally throughout the generations.

In the years since our initial publication~\cite{us1}, we have become aware of excellent work taking place around the world. 
This includes traditional approaches to comparative mythology, folktales and to epic literature.
Significant amounts of data have been gathered and we think that these are amenable to  mathematical, statistical and computational methodologies. 
Indeed,  fascinating   approaches include 
conceptual mapping in a large folklore corpus \cite{TT};
analyses of virtual worlds \cite{OM}; 
studies of cognitive constraints in literature \cite{RD};
principal component analyses of  distributions of folklore motifs \cite{YB};
and phylogenetic approaches to folk tales \cite{JdH,JT}. 
To deepen and broaden quantitative investigations into mythology, annals and folktales, more collaboration between the humanities and sciences is needed.
We have hosted interdisciplinary workshops which include physicists, applied mathematicians, complexity theorists, computer scientists, anthropologists, psychologists, authors, film makers, artists, historians, medievalists, mythologists and other humanities scholars. 
Resulting from these, an edited volume is under preparation and is scheduled to be published in 2016 \cite{MMM}.
We hope that these activities will spur further collaborative work and help break down the barriers that are perceived to exist between our two cultures.

%%%%%%%%%%%%%%%%%%%%%%%%%%%%%%%%%%%%%%%%%%%%%%%%%%%%%%%%%%%%%%%%%%%
\ack
%%%%%%%%%%%%%%%%%%%%%%%%%%%%%%%%%%%%%%%%%%%%%%%%%%%%%%%%%%%%%%%%%%%

%This work was  supported by the College for the Statistical Physics of Complex Systems, Leipzig-Lorraine-Lviv-Coventry $({\mathbb L}^4)$ and by The Leverhulme Trust under grant number F/00732/I.
We would like to thanks Peter Berresford Ellis for helpful comments on the text. 
We also thank staff at the Rathcroghan Visitor Centre for hospitality at a conference at Cruachan A{\'{i}}.

%%%%%%%%%%%%%%%%%%%%%%%%%%%%%%%%%%%%%%%%%%%%%%%%%%%%%%%%%%%%%%%%%%%

\end{document}